\documentclass[sigplan,screen,preprint]{acmart}

\usepackage[utf8]{inputenc}
\usepackage{pmboxdraw}
\usepackage{pifont}
\usepackage{newunicodechar}
\newunicodechar{✓}{\ding{51}}
\newunicodechar{✗}{\ding{55}}

\AtBeginDocument{%
  }

\setcopyright{acmlicensed}
\copyrightyear{2025}
\acmYear{2025}
\acmDOI{XXXXXXX.XXXXXXX}
\acmConference[ICFP '25]{The ACM SIGPLAN International Conference on Functional Programming}{Oct 12--18, 2025}{Singapore}
\acmISBN{978-1-4503-XXXX-X/2018/06}

\begin{document}

\title{Highly Interactive Testing for Uninterrupted Development Flow}

% Possible Titles:

% - Deceptively Simple Interactive Testing Library
% - Shortening Feedback Loop for Tests
% - TRDD: Test-REPL Driven Development
% - Highly Interactive Testing Library
% - Improving ergonomics of Test Driven Development
% - Unbreaking Development Flow with Interactive Tests.
% - Rerunning Tests without Reexecuting the Program.
% - Improving DX with Interactive Tests.
% - Improving DX for Tests and with Tests
% - Interactive Testing: The Workflows and The Implementation
% - Interactive Testing: The Workflows and Implementation Behind Uninterrupted Development Flow
% - Interactive Testing: Unbreaking Development Flow for Highly Interactive Environments
% - Highly Interactive Testing for Uninterrupted Development Flow

\author{Andrew Tropin}
\email{andrew@trop.in}
\affiliation{%
  \institution{Departament of Blog Posts}
  \country{World}
}

\renewcommand{\shortauthors}{Tropin et al.}

\begin{abstract}
  Highly interactive development environments (HIDEs) enable
  uninterrupted development flow through continuous program evolution
  and rapid hypothesis checking. However, traditional testing
  approaches---typically executed separately via CLI---isolate tests
  from HIDE tooling (interactive debuggers, value and stack
  inspectors, etc.) and introduce disruptive delays due to coarse
  execution granularity and lack of runtime context. This disconnect
  breaks development flow by exceeding critical attention
  thresholds. In this paper we present a library that provides runtime
  representation for tests, allowing tight integration with HIDEs, and
  enabling immediate access to HIDE tooling in the context of test
  failure. We then describe development workflows enhanced with
  testing and demonstrate how they achieve subsecond test reexecution
  times crucial for maintaining developer focus.
\end{abstract}




\received{24 June 2025}
\received[revised]{12 March 2009}
\received[accepted]{5 June 2009}

\maketitle

\section{Introduction}
Highly Interactive Development Environments (HIDEs) like Lisp
Machines, Jupyter Notebooks, REPLs, and browser Devtools provide live
systems where developers dynamically update functions, modify
variables, and evaluate expressions without restarting processes or
losing the state.

HIDEs enable uninterrupted development workflows constrained by
human's cognitive capabilities rather than system's latency. To
achieve this, all interactions---code reloading, value inspection,
debugging---must respond within strict time thresholds (detailed in
\autoref{sec:idea}), preserving attention.

Imagine working on a project using a HIDE\footnote{This includes text
  editors/IDEs integrated with a live process via specialized
  protocols like Jupyter\cite{perezMessagingJupyterJupyter2015},
  nREPL\cite{emerickNREPLAsyncronousProtocol2010}, or stdin/stdout of
  the usual REPL.}: you hot-reload code into the running program and
immediately see results\footnote{Widely popular workflow in scientific
  research (e.g. Jupyter notebooks) and general programming
  (e.g. Clojure, Common Lisp)}. This enables rapid experimentation
through continuous hypothesis checking and fast iteration. However,
when running tests, you reach for the terminal, type \texttt{make
  check}, and wait for dozens of minutes. In medium-sized projects,
this takes \textasciitilde10 minutes (\autoref{sec:problems}).

Crucially, you lose access to all interactive development tools
(\autoref{sec:problems}): debuggers, disassemblers, and value/stack
inspectors \cite{wingoInteractiveDebuggingGuile2010}
\cite{batsovCIDERValueInspector2014}
\cite{tropinActuallyUsefulStack2025}. Moreover, test results are
dumped to unstructured text logs, forcing manual inspection via
\texttt{grep}/\texttt{less}/\texttt{awk} while toggling between
terminal and IDE. This disconnect creates significant friction in the
development workflow and has at least two negative consequences:

\begin{itemize}
\item Developers avoid writing/utilizing tests due to perceived workflow penalties
\item Test execution creates disruptive context switches when attempted
\end{itemize}

Existing optimization techniques (faster recompilation, parallel test
execution) still fail to achieve subsecond feedback cycles, and
cooperation of HIDE tooling with tests.

% Why existing solutions doesn't solve the problem

We present a method of gradual integration of testing into
HIDE/REPL-based development approaches
(\autoref{sec:workflow-interactive-testing}), and introduce
development workflows enabling test executions and reruns in
sub-second interals (\autoref{sec:workflow-fail-fast}) with immediate
access to HIDE tooling (\autoref{sec:workflow-throw-on-failure}). Our
primary contributions are:
\begin{itemize}
\item A runtime-centric library design with a minimal API that
  integrates seamlessly with highly interactive development
  environments (\autoref{sec:defining}, \autoref{sec:running}).
\item Practical workflows that enchance HIDE-based development
  approaches with testing, and provide radical improvement of test
  execution times (\autoref{sec:workflows}).
\end{itemize}

While the concepts described in the paper are broadly applicable
across programming languages, this work focuses on implementing and
evaluating them within the Scheme ecosystem.

% Crucially, unlike incremental test runners, our approach preserves
% full runtime context and debuggability during test execution.

\section{The Problems and The Idea}
\label{sec:idea}
\label{sec:problems}
When the time between intent and accomplishing an action exceeds
certain thresholds, the development flow breaks.  This occurs whether
it takes too long to obtain a test execution results or get a local
variable's value after an exception has been thrown.

The thresholds are determined by human psychological
factors\cite{millerResponseTimeMancomputer1968} and are the
following\cite{nielsenResponseTimeLimits1993}:
\begin{itemize}
\item \textbf{0.1 seconds (Instantaneous)}: Feels as immediate
  response - no special feedback or treatment needed.
\item \textbf{1.0 seconds (Uninterrupted Flow)}: The limit for
  maintaining continuous flow of thought - developer notices delay but
  flow remains unbroken.
\item \textbf{10 seconds (Attention Lost)}: The attention limit -
  beyond which developer shifts focus and suffers significant
  context-switching penalties.
\end{itemize}

The primary idea is to reduce test execution times to under one
second\footnote{There are cases, when it's not possible to achive one
  second intervals, but we provide workarounds.} and provide rapid
acces to the usual HIDE's development tooling (interactive debugger,
goto definition/place of exception, value inspector, etc).  We achieve
this by integrating tests into highly interactive development
environments and using corresponding workflows
(\autoref{sec:workflows}), but before discussing them, let's take a
look at usual Scheme project testing approach and its limitations.

As an example, consider GNU
Guix\cite{courtesFunctionalPackageManagement2013}, a medium-to-large
Scheme project, a functional package manager and operating system
implemented in Guile. It has a good test coverage and it exemplifies
the most widespread testing approach in the Scheme ecosystem, an
approach found in projects ranging from small
libraries\cite{herdtSchemejsonrpcTestsClientscm2021} to distributed
object programming
environments\cite{christinelemmer-webberGoblinsTestsTestcorescm2018}.
This testing approach combines usage of
SRFI-64\cite{bothnerSRFI64Scheme2006} library with CLI test runners
and text logs. Let's take a quick glance at how this testing workflow
looks and feels.

While Guix can be developed with HIDEs like
Ares\cite{tropinAresAsyncronousReliable2023},
Geiser\cite{ruizTopGeiserUser2009} or just a basic REPL and provide
uninterrupted development flow experience, the tests are completely
disconnected from those tools and are executed from CLI.  The usual
\texttt{make check} takes around 10 minutes on modern
hardware\footnote{Intel's i7-1260P CPU} with all the optimizations and
parallelization.

\begin{verbatim}
PASS: tests/accounts.scm
PASS: tests/base16.scm
...
PASS: tests/packages.scm
FAIL: tests/pack.scm
...
# TOTAL: 2604
# PASS:  2569
# SKIP:  24
# XFAIL: 6
# FAIL:  5
...
See ./test-suite.log
...
Exited with code 2 at Thu Jul 3 after 08:46:11
\end{verbatim}

While specific test files can be executed individually (e.g.,
\texttt{make check TESTS=test/pack.scm}), this approach lacks support
for parallel test execution. And even when targeting single files,
execution times remain high—ranging from dozens of seconds to minutes—
which significantly exceeds our subsecond target.

Execution time is not the only thing introducing delays and thus
interruptions to the flow here.  Testing results are available as text
files mixing output of the program and test logs\footnote{We could
  improve here and made them well-structured and easily parseable, but
  it won't solve all the issues and most of the following arguments
  still applies}.

There is no integration between test results and development
environment and no way to immediately perform desired action on them
(go to the source location of the failed test, see the state of the
stack at the moment of failure or evaluate expression in the context
of test's environment).  In the best case each of those actions
require manual intervention and, more importantly, time and attention
of the developer, in the worst they are impossible to achieve, because
the runtime information and state is not available anymore.

For example, to get a current stack or other runtime clue it will
require to setup the similiar environment as test code has and
evaluate assert expression in it.  It can be done with HIDE and some
manual work, but it far exceeds even 10 seconds time frames. It
creates a lot of friction and make it less likely people use the tests
often or at all.

Let's recap the hurdles of the current appoarch:

\begin{itemize}
\item Tests execution is not granular, it impossible to quickly select
  only related tests. This leads to huge delays from long test runs.
\item Tests doesn't have runtime representation, so it is hard to
  rerun only failed tests or run tests in parallel\footnote{While with
    SRFI-64 it's possible to run separate files with tests in
    parallel, the tests inside the file are executed sequentially}.
\item Testing results are in text files and time from observing the
  result to the action is more than dozen of seconds.  When test
  fails, you can't immediately access debugger, value/stack inspector,
  and other tools.
\end{itemize}

Now, when we have a good understanding of areas we can improve, let's
see how we can do so.

% TODO: [Andrew Tropin, 2025-07-12] Highlight a bit related works or
% forward reference the related works section.

HIDEs/REPLs already enable evaluating arbitrary code fragments from
modules to individual expressions, allowing live modifications to
propagate quickly (typically within milliseconds).  We adopt this same
paradigm for testing. In addition to that, we delay assertion
execution by creating runtime entities and attaching assertion
thunks\footnote{A function with zero arguments} to them (Section
\ref{sec:defining}) and only after that run those thunks.

It gives us an execution granularity and allows to maintain runtime
information about tests.  Also, we get a seamless access to all the
development tooling of HIDE.  While we explore implementation
specifics in Sections~\ref{sec:defining} and~\ref{sec:running}, this
fundamental shift already narrows the testing-development gap and
already eliminates many of the workflow interruptions by just treating
tests as first-class runtime entities.

Further, we build workflows and HIDE integrations upon that
functionality.  We start by cutting unecessary delays and frictions of
usual HIDE/REPL-driven development workflow by gradually substituting
visual inspection of evaluation results with automated test-based
checking (\ref{sec:workflow-interactive-testing}).  This allows to
make testing a part of HIDE-driven workflow and facilitates systematic
accumulation test suites (\ref{sec:workflow-more-tests}).

Once the number of tests in the project reaches a critical mass, their
execution delays begin to disrupt the development flow.  To address
this challenge, we leverage the tests' runtime representation and
information about previous runs to implement two key mechaniques:

\begin{itemize}
\item Interactive narrowing of test suites
  (\ref{sec:workflow-selecting}). It allows to run only most relevant
  tests for the current context.
\item Rerunning of only failing tests
  (\ref{sec:workflow-rerun-failed}) and bringing result as soon as
  failure occurs (\ref{sec:workflow-fail-fast}).
\end{itemize}

It allows to reach sub-second goals for test re-execution times, but
we don't stop here.  The last gap remains between test failure
occurrences and developers obtaining diagnostic insights.  We
eliminate this last discontinuity between test executions and HIDE
tooling by bringing the interactive stack inspector and debugger into
the test failure context (\ref{sec:workflow-throw-on-failure}).

The next two section discuss the design and implementation of testing
library, followed by a section demonstrating how to utilize the
library and its integration into HIDE to achieve declared objectives.

\section{Defining Tests}
\label{sec:defining}
The primary goal of the test defining API is to provide runtime
entities representing tests and test suites, so they can be used
interactively from HIDE.  The test loading and execution is
intentionally shifted to the test runner, so developer can adjust
these behaviors to suite their spicific or our suggested
\nameref{sec:workflows}, more on that in \nameref{sec:running}.  The
default test runner is tailored for interactive bottom-up development
approach with minimal to no special support from HIDE and this section
reflects it.  However, keep in mind that it's not set in the stone and
it can be and will be different, when we start exploring more advanced
testing workflows.

\subsection{Assertion}
To demonstrate and explain the test defining API, we build a test
suite using a bottom-up approach. To start doing so, we need an
assertion mechanism—the most frequently used and fundamental part of
every testing framework. As for the name, \texttt{is} is short and
concise, and we can use it like this:

\begin{verbatim}
(is (= 4 (+ 2 2)) ;; => #t
(is 'hello) ;; => 'hello
(is (any even? (list var1 var2 var3))))
(is (throws-exception? (f 1 2)))
(is (exception-message=? (f 1 2) "hi"))
\end{verbatim}

We can direcly execute those assertions using a HIDE or just a REPL.
With the default test runner it will immediately display return values
and we can understand if our expectations are met.  In addition to
that, a test reporter \ref{sec:test-reporters} can provide other
information: execution time, thrown exceptions, values of evaluated
arguments and anything else one can find handy for their use case.
More advanced behaviors of test runner are covered in
\nameref{sec:test-runner} and \nameref{sec:workflows}.

Implementation-wise, \texttt{is} is a macro (or a similar
construct\cite{shuttFexprsBasisLisp2010}) to achieve two key capabilities:
capturing the code under assertion verbatim and delaying its execution
\cite{marshallJRMsSyntaxrulesPrimer2013}
\cite{nieper-wisskirchenExtendingLanguageWriting2023}. The first
capability enables more detailed test reports by preserving
expressions verbatim, while the second allows evaluation in a
controlled environment to handle exceptions, validate return values,
or perform other operations set by test runner. For example, when
evaluating \texttt{(is (equal? 'hi 'hey))}, a test report can yield a
descriptive message:

\begin{verbatim}
In expression (equal? 'hi 'hey):
  `hi` and `hey` are not `equal?`
\end{verbatim}

As for safe and controlled execution of assertions, we won't be
deciding on that behavior now; we delegate it to the
\nameref{sec:test-runner}. This will make our core API stable, while
the library behavior remains highly customizable. In fact, this design
decision makes the library so flexible that we can introduce new
functionality into assertions on the test runner side without
modifying the library itself. For example, should one want to annotate
an expression under assertion with a string, we would be able to
implement it like this.

\begin{verbatim}
(is (with-description (= 4 (+ 2 2)) "school math"))
\end{verbatim}

We can unwrap \texttt{(with-description (= 4 (+ 2 2)) "school math")}
on the test runner side, attach the \texttt{"school math"} annotation
to it, and, for example, make a test reporter print \texttt{"Checking
  school math"} before executing the expression.  To improve
\texttt{with-description}'s usability we could use a syntax parameter
\cite{barzilayKeepingItClean2011}, but unfortunately macros don't compose
well and we can't delay syntax parameterization to the runtime and do
it in a test runner.  This is where the library could benefit from
F-exprs\cite{shuttFexprsBasisLisp2010}.

One more point to note: since the test runner controls assertion
execution, it defines the possible assertion outcomes. The default
implementation supports three outcomes:
\begin{itemize}
\item \textbf{Pass}: When the expression returns a truthy value.
\item \textbf{Failure}: When the expression returns a falsy value.
\item \textbf{Error}: When an exception is raised during execution.
\end{itemize}
We will use these knowledge in Section \ref{sec:test}.

To summarize the most crucial aspects of the implementation:
the \texttt{is} macro captures an unevaluated expression and creates:
\begin{enumerate}
\item An assertion body thunk (to run the expression).
\item An arguments thunk (to evaluate and list arguments).
\end{enumerate}
These are passed to the test runner for processing. The default runner:
\begin{enumerate}
\item Immediately executes the assertion.
\item Invokes a test reporter.
\item Returns the expression's value or re-raises any exceptions.
\end{enumerate}

% We didn't mention arguments thunk earlier, but it's needed

\subsection{Test}
\label{sec:test}
While individual assertions are valuable, real-world programs often
require verifying invariants through multiple related assertions. We
introduce a \texttt{test} entity to group these assertions for
combined execution. Its syntax is straightforward:

\begin{verbatim}
(test "Description of the test"
  (is #t)
  (is (= 5 (+ 2 2)))
  (is 'hello))
\end{verbatim}

Unlike the \texttt{is} macro, every \texttt{test} includes a mandatory
description. This is necessary because tests represent more complex
units of verification than individual assertions, requiring explicit
context to identify their purpose. The description serves both
reporting needs and test hierarchy inspection
(Section~\ref{sec:test-suite}). Implementation-wise, \texttt{test}
wraps its expressions in a thunk (zero-argument procedure), attaches
metadata, and registers it with the test runner.

The implementation (Section \ref{implementation}) of the \texttt{test}
macro allows controlling nearly all aspects of its behavior via the
test runner. However, by default, when a test is evaluated standalone,
the runner immediately executes test's body thunk, signals reporters
to relay execution progress, and returns \texttt{\#<unspecified>}
value.  The example below demonstrates the output of the test runner:

\begin{verbatim}
┌Test Description of the test
#t
✓
(= 5 (+ 2 2))
✗ 5 and 4 are not =
'hello
✓
└Test Description of the test
\end{verbatim}

The \texttt{\#<unspecified>} value prevents unintentional use of a
return value of \texttt{(test ...)} form, avoiding coupling with the
surrounding environment.  Though the default runner executes test's
body immediately, advanced workflows like parallel execution
(\ref{sec:workflow-parallel}) and dynamic test suites
(\ref{sec:workflow-selecting}) delay execution—only registering tests
for later runs, so results aren't available at the definition site.

This deferred execution enables constructing the test hierarchy
upfront, providing the ability to preview and introspect the test
suite structure. This allows for flexible execution: running tests in
parallel, random order, selectively rerunning, or performing other
kind of adavnced interactions. Consequently, test bodies should remain
as self-contained as possible, minimizing dependencies on execution
order or surrounding environment.

The default test runner prohibits nested tests, as it violates the
invariant that tests solely contain assertions. When encountered, it
throws an invalid-nesting exception. This behavior can be altered by
implementing a custom test runner. Alternatively, users can
dynamically override the current test runner using scoped parameters
\cite{feeleySRFI39Parameter2003} to permit nested test definitions,
though we defer full discussion to Section \ref{sec:test-runner}

Last but not least, we introduce the ability to attach metadata to
tests. This metadata provides granular control over test execution
behavior - we can skip tests, implement 'expected to fail'
functionality, or enable other specialized behaviors. The syntax
remains straightforward:

\begin{verbatim}
(test "Description of the test"
  #:metadata '((key1 . val1) (key2 . val2))
  (is #t)
  (is (= 5 (+ 2 2)))
  (is 'hello))
\end{verbatim}

We demonstrate application scenarios for this metadata in Section
\ref{sec:test-runner}. Before proceeding to execution details, let us
summarize key characteristics of our test definition model:

\begin{itemize}
\item The \textit{test} construct groups related assertions while
  prohibiting nested test definitions.
\item Test definitions return \texttt{\#<unspecified>} to avoid
environmental coupling through return value dependencies.
\end{itemize}

This design enforces test atomicity, requiring operational
independence to support out-of-order execution, parallelization, and
other workflows.  The default test runner executes isolated tests
immediately while queueing suite-contained tests for scheduled
execution.

\subsection{Test Suite}
\label{sec:test-suite}
In the previous section we noted that tests cannot be nested. This is
an intentional design choice, we want test functionality to remain
explicit and focused: tests are specifically responsible for
collocating related code, asserts, and respective metadata. Test
suites, on the other hand, handle organization of tests and
construction of test hierarchies. Syntactically, test suite definition
closely resembles test definition:

\begin{verbatim}
(test-suite "Simple test suite"
  #:metadata '((optional . meta) (data . alist))
  (test "Simple test 0"
    (is #t))
  (test "Simple test 1"
    (is (= 5 (+ 2 2)))
    (is 'hello)))
\end{verbatim}

Unlike tests, test suites support nesting—they can contain other test
suites, individual tests, or a combination of both—but they cannot
contain assertions directly. This nesting constraint can be altered by
customizing the test runner.

Evaluation behavior also differs slightly: when a test suite is
evaluated, it triggers the test runner to execute the suite's
body. Upon encountering tests, the runner registers their bodies
thunks for later execution rather than evaluating them
immediately. Once the entire test hierarchy is constructed, by
default, the test runner traverses it and executes the tests.

When organizing and persisting test suites in modules, we generally
want to avoid executing them during module loading, as running
arbitrary computations with potential side effects is an antipattern
and overall questionable practice. Therefore, for reusable test
suites, we must delay their execution. We achieve this as follows:

\begin{verbatim}
(define sample-tests
  (test-suite-thunk "sample-tests"
    (test "Test 1"
      (is #t))
    (test "Test 2"
     (is (= 5 (+ 2 2)))
     (is 'hello))))
\end{verbatim}

Note that we use \texttt{test-suite-thunk} instead of
\texttt{test-suite} here. This can be equivalently achieved using the
dedicated \texttt{define-test-suite} macro as shown below:

\begin{verbatim}
(define-test-suite sample-tests
  (test "Test 1"
    (is #t))
  (test "Test 2"
   (is (= 5 (+ 2 2)))
   (is 'hello)))
\end{verbatim}

Test suites defined this way can be executed later in a REPL/HIDE
by evaluating \texttt{(sample-tests)}, since \texttt{sample-tests} is
a standard zero-argument procedure. While test suites can be defined
in any module, Section~\ref{sec:discovery} provides recommendations
for organizing tests and module structures.

To get a better understanding of how test suites work, let's extend our sample test suite with additional nesting and run it.

\begin{verbatim}
(define-test-suite sample-tests
  (test "Test 1"
    (is #t))
  (test-suite "Nested test suite"
    (test "Test 2"
      (is (= 5 (+ 2 2)))
      (is 'hello))))
(sample-tests)
\end{verbatim}

The output of a default test runner equipped with a verbose and
hierarchical reporters (Section~\ref{sec:test-reporters}) appears as
follows:

\begin{verbatim}
┌> sample-tests
| + test Test 1
|┌> Nested test suite
|| + test Test 2
|└> Nested test suite
└> sample-tests

┌Test Test 1
#t
✓
└Test Test 1

┌Test Test 2
(= 5 (+ 2 2))
✗ 5 and 4 are not =
'hello
✓
└Test Test 2
((errors . 0)
 (failures . 1)
 (assertions . 3)
 (tests . 2))
\end{verbatim}

One implementation note regarding test suite representation: we
construct them as procedures augmented with a \texttt{(test-suite?
  . \#t)} flag, crucial for the discovery mechanism discussed in
\nameref{sec:discovery}.  While Guile Scheme's procedure properties
feature provides a natural storage mechanism for this flag,
alternative implementations could employ a separate registry data
structure.

A small summary. Test suites are used to organize tests and build
hierarchies. Test suites can contain tests and other test suites, but
not assertions. Nesting behavior is defined by a test runner. To
prevent immediate evaluation of top-level test suites on module
loading, one should use \texttt{test-suite-thunk} or
\texttt{define-test-suite}.

% \subsection{Reloading Tests and Suites}
% This is area, where the library can be improved.  For now default test runner just resets the registry with the last loaded test suite.

% We can associate a test suite with a variable, so when the variable is
% updated, we update corressponding test suite entry inside test
% runner's registery.

% We can make a separate test runner, which just collects, filters and
% organaizes tests.

% Rename test runner to test manager?

\section{Running Tests}
\label{sec:running}
\label{sec:test-runner}
While the previous section demonstrated how assertions, tests, and
test suites and the default test runner already enable interactive use
in HIDEs/REPLs for core functionality, significant potential remains
to enhance test usability and workflows. This section examines core
test runner's APIs and execution mechanics, and delivers practical
tools for constructing interactive testing workflows.

The test runner handles all the heavy lifting within the library. As
shown in the appendix \ref{implementation}, the macro API remains
deliberately thin, with all core logic implemented in the test
runner—including enforcement of test and test suite nesting
invariants, test execution logic, assert results handling, and
more.  This design offers significant flexibility: users maintain a
unified API for test definition while gaining full control over
execution behavior through custom runners.

The test runner is stateful and typically long-lived object, created
once and usable for hours. Implementation-wise, it's a single-argument
function (handling messages) - more precisely, a closure over stateful
atomic variables. The only interaction method is calling this function
with specific messages, but before exploring those capabilities, we'll
address methods to create and access it.

\subsection{Accessing Test Runner}
\label{sec:accessing-test-runner}
To create a test runner one simply calls a function returning a test
runner object:

\begin{verbatim}
(make-suitbl-test-runner
  #:test-reporter test-reporter-silent
  #:parallel-execution? #f)
\end{verbatim}

The test runner is connected to the test defining API (described in
Section \ref{sec:defining}) via a dynamically scoped variable
\cite{feeleySRFI39Parameter2003}, \texttt{test-runner*}. All test
defining APIs interact with this variable, enabling the test runner to
be quickly changed when needed, thereby altering the behavior of
\texttt{is}, \texttt{test}, and \texttt{test-suite}.

\begin{verbatim}
;; set test runner globally
(test-runner* new-test-runner)

;; set test runner temporary
;; for the extent of parameterize
(parameterize ((test-runner* new-test-runner))
  ...)
\end{verbatim}

This approach enables creating special macros or functions that
temporarily alter the current test runner. We demonstrate this with
the following practical example used when testing the library and test
runner implementation themselves\cite{TestEnvironmentMacro2025}:

\begin{verbatim}
(define-syntax test-environment-silent
  (lambda (stx)
    (syntax-case stx ()
      ((_ body body* ...)
       #'(parameterize
          ((test-runner*
            (make-suitbl-test-runner
             #:test-reporter
             test-reporter-silent)))
           body body* ...)))))
\end{verbatim}

\subsection{Loading Tests}
\label{sec:loading-tests}
After determining how to access the current test runner, we now
examine its capabilities. Loading tests prior to execution serves
three key purposes:

\begin{enumerate}
\item Understanding the test hierarchy structure
\item Enabling dynamic scheduling of tests
\item Preserving state information between runs
\end{enumerate}

The easiest way to load a test suite is to call a function created
with \texttt{test-suite-thunk} macro.  Considering example from
\nameref{sec:defining} section, we can load it like this
\texttt{(sample-tests)}.  The more explicit way will look like this:

\begin{verbatim}
((test-runner*) ; get current test runner
 ;; call run-test-suite-body-thunk method
 `((type . run-test-suite-body-thunk)
   (test-suite-body-thunk .
    ,(procedury-property
      sample-tests 'test-suite-body-thunk))))
\end{verbatim}

Both approaches are functionally equivalent; the latter simply
demonstrates what occurs internally when calling
\texttt{sample-tests}. The \texttt{run-test-suite-body-thunk} method
initializes required state variables and executes the body of the
\texttt{(test-suite description body ...)} definition. This triggers
evaluation of nested test suites and test definitions, with each
subsequently invoking the current test runner to process their
respective \texttt{test-suite-body-thunk} or \texttt{test-body-thunk}.
The fact that tests and test suites wrap their bodies into thunks and
pass them to test runner instead of immediately executing them is what
allows us to load tests and build the test hierarchy.

After all tests are loaded we get a nested data structure, which looks
like this:

\begin{verbatim}
("suite: base-test-runner-tests"
 ("suite: is-usage-tests"
  "test: basic atomic values"
  "test: predicates"
  "test: is on its own in empty env"
  "test: nested is and is return value")
 ("suite: test-macro-usage-tests"
  "test: simple test case marked as slow"
  "test: zero asserts test macro works fine"
  "test: standalone test macro usage")
 ("suite: test-suite-usage-tests"
  ("suite: test suite with metadata"
   "test: simple")
  ("suite: nested-suites-and-test-macros-tests"
   "test: nested test macro usage is forbidden"
   "test: test-suite nested in test is forbidden"
   ("suite: nested test suite 1"
    "test: test macro 1#1"
    ("suite: even more nested test suite 1.1"
     "test: test macro 1.1#1")))))
\end{verbatim}

For the readability we use strings, but in fact the data structure
contains corresponding thunks with attached metadata. The list
represents a test suite, the first element contains
test-suite-body-thunk, the rest elements are test-body-thunk and lists
(nested test suites).

The default test runner after building the test hierarchy will
immediately start traversing this data structure and executing tests;
to avoid this behavior, one needs to add \texttt{(execute? . \#f)} to
the message sent to the test runner. More information about running
options appears in \autoref{sec:running-options}.

\subsection{Discovery}
\label{sec:discovery}
% TODO: [Andrew Tropin, 2025-07-18] Proofread and fix wording

Defining tests and loading them are covered, but the persisting is not
yet.  This section describes the project source code organization
approach that helps dealing with tests more comfortable and API, which
helps loading whole project and module's test suites.

Tests and suites are usual Scheme code, so we will use the same
organizing principle: we will store them in modules.  We suggest
storing test in separate modules and even in separate
\texttt{\%load-path} \cite{jafferLoadPathsGuile1997}.  Here is an
example:

\begin{verbatim}
;; (my-project tool) module
src/guile/my-project/tool.scm

;; (my-project tool-test) module
test/guile/my-project/tool-test.scm

\end{verbatim}

This approach enables us to do two thing:
\begin{enumerate}
\item Make sure test's code excluded from production build by
  elminating test/ from \texttt{\%load-path}.
\item Have a clear correspondence between the primary module and test
  module, so we can easily discover most related tests for the current
  module with \texttt{get-test-module} function.
\end{enumerate}

\begin{verbatim}

(get-test-module '(ares suitbl))
;; => #<directory (ares suitbl-test) 7f7994629960>

(get-module-test-suites
 (get-test-module '(ares suitbl)))
;; (list test-suite1 test-suite2 ...)

(get-all-test-modules)
;; =>
;; (#<directory (integration-test) 7f79dc267f00>
;;  #<directory (ares evaluation-test) 7f79704633c0>
;;  ...
;;  #<directory (ares suitbl-test) 7f7994629960>)

\end{verbatim}

With this API we are able to load arbitrary test suites available in
our project easily.

% \subsection{Scheduling Tests, Building Dynamic Test Suites}
% \label{sec:scheduling}

% TODO: [Andrew Tropin, 2025-07-10] Provide an interface for manually
% picking tests.

% TODO: [Andrew Tropin, 2025-07-10] Saves test runs, so we have a test
% run history and can rerun previous test suites.

% TODO: [Andrew Tropin, 2025-07-10] Load tests should save timestamps,
% so the tests with same names can be deffirentiated.

% TODO: [Andrew Tropin, 2025-07-10] The loaded test should contain the
% hierarchy information attached to it.

% We want the whole test hierarchy to be available via consult
% interface, so we can interactively filter it by description/metadata
% and create dynamic test suites out of the filtered list.  Or maybe
% just a separate buffer, with possibility to mark multiple entries
% and consult interface for multi-marking?

% CSS/jq selectors for making tests!

% !!!!!!!!!!!!!!!!!!!!!!!!!!!!!!!!!!!!

% It solves another annoying problem: multiple re-evaluation or
% evaluation history.  You don't actually need the evaluation history,
% you can just dynamically decide which pieces of code to re-evaluate.

\subsection{Running Options}
\label{sec:running-options}
% TODO: [Andrew Tropin, 2025-07-20] Cleanup this section

% Disclaimer: This is WIP and not all of this functionality available at
% the library yet.

Now, when we have all necessary tests loaded into the test runner, it's time to run them.

For this we will be using HIDE's interface to the test runner.  It's
implemented as a transient menu for Emacs and allows to set execution
options.  By default tests are executed in the random order
sequentially as one big test suite, but the following flags allow to
adjust this behavior.

\begin{itemize}
\item \textbf{Preserve hierarchy}: group by test suite, and build
  hierarchy.
\item \textbf{Sequential run}: preserve order in which tests were
  loaded.
\item \textbf{Parallel run}: executes test in parallel.

\item \textbf{Fail fast}: short circuit, stop on the first failure.
\item \textbf{Failing first}: run previously failed tests first.
\item \textbf{Debug on failure}: throw the exception and let HIDE
  handle it.
\item \textbf{Rerun failed}: Rerun failed test (unless there are
  failing tests, if there is none, run the whole test suite).
\item \textbf{Quick filters}: add a filter to reduce number of test.
  to execute.
\end{itemize}

Those options allow to radically cut the test execution
time. \autoref{sec:workflows} explains how.

\subsection{Test Reporters, JUnit and TAP}
\label{sec:test-reporters}
Test reporters are simple stateless functions, which can accept one
argument (message).  This message contains the type of the event and
all related information, it also contains an atomic-box object created
inside test runner and this object can be used to store stateful data.
When test reporter encounter event it can handle it uses
\texttt{test-reporter-output-port*} to print report to it.  If test
reporter can't handle this event type it return \texttt{\#f},
otherwise it returns non-falsy value.  It allows to combine multiple
reporters with \texttt{test-reporters-use-all} and
\texttt{test-reporters-use-first} functions.

Default test runner uses only one test reporter and sends all the
related events to it.  However, with functions for combining test
reporters mentioned above or just arbitrary functions, one can
construct any reporting behavior.

There are a few simple reporters implemented at the moment.

\begin{verbatim}
test-reporter-silent
test-reporter-logging
test-reporter-unhandled
test-reporter-hierarchy
test-reporter-verbose
test-reporter-dots
\end{verbatim}

And a few compound\footnote{A compound test reporter is a reporter
  made from a combination of other reporters}, we will just take a
look at test-reporter-base, because it demonstrates the usage of test
reporter combinators, other compound test reporters are similar.

\begin{verbatim}
(define test-reporter-base
  (chain (list test-reporter-verbose test-reporter-hierarchy)
    (test-reporters-use-all _)
    (list _ test-reporter-unhandled)
    (test-reporters-use-first _)))
\end{verbatim}

\section{Workflows}
\label{sec:workflows}

This is the place, where we save human lifes or at least their
precious time and attention.  Throughout this section, we present
various workflows designed to eliminate friction and delays while
incrementally establishing more efficient software development
practices. These workflows are primarily composable, enabling flexible
mixing, matching, and layering.

Our goal is to reduce turnaround times to sub-second intervals to
maintain uninterrupted development flow. We acknowledge scenarios
where this target may not be achievable (e.g., when expression
evaluation inherently exceeds one second or when whole test suite must
be re-executed) and provide corresponding mitigation strategies.

We begin by integrating testing into standard HIDE/REPL-driven
development workflows. Subsequently, tests created during this phase
are systematically persisted within the project. Once a critical mass
of tests is established, we introduce specialized workflows that
execute them without incurring disruptive delays or interrupting the
development flow.

Furthermore, to quantify common performance characteristics: the
majority of tests in typical projects execute
near-instantaneously. Based on empirical observations, approximately
10\% run within one second, while less than 3\% require longer
execution times. Development workflows predominantly interface with
tests in these first two categories, with suites typically comprising
between zero and a few dozen tests during active development
sessions. \autoref{sec:workflow-selecting} and
\autoref{sec:workflow-fail-fast} elaborates on the mechanisms allowing
to work with only related to the current context narrower test suites.

\subsection{Interactive Bottom-up Testing}
\label{sec:workflow-interactive-testing}
\begin{itemize}
\item \textbf{Before}: 20-50 seconds (to evaluate a few expressions
  and visually inspect them).
\item \textbf{After}: 0.5-10 seconds (to evaluate the previous test
  expression).
\end{itemize}

In HIDE/REPL-driven development, we construct complex systems from
small, modular components—functions, expressions, or discrete
values. When properly designed, these elements require minimal setup
for execution and can be independently evaluated to verify intended
outputs. This approach inherently supports rapid iteration cycles.

However, we sometimes find ourselves repeatedly reevaluating the same
expression and manually verifying its return value—a time-consuming
practice. This scenario presents an opportunity for interactive
testing: rather than executing the expression and inspecting the
result, we instead write an assertion that automatically verifies the
output.

We are rarely interested in only a single property of the entity we
develop, so we need multiple assertions to check if all the desired
invariants hold.  Manual reevaluation of individual assertions becomes
impractical as their quantity increases.  Fortunately, we have a test
entity described in Section~\ref{sec:test} that allows to group
multiple asserts and run them all at once.  In HIDE one can do it by
navigating to the place, where the test body is written and pressing a
hotkey (C-c C-t C-c).  Moving to the place where expression of the
test is located usually requires several seconds, which already quite
a lot according to our requirements, so we optimize it even further
with \nameref{sec:workflow-rerun} workflow.

To sum up: when frictions and delays from continious reevaluation and
visual inspections start to appear, this approach allows to gradually
introduce assertions and tests into our development flow, thus
removing those obstacles and keeping the flow intact.

\subsubsection{Tests as an Evaluation History}
However, there are concerns about the usefulness of the tests for
HIDE-driven development in some use cases:

\begin{quote}
On the smaller sub-function scale, writing tests would be a waste of
time, though. Tests are also no option when you work on something
visual, like UI\cite{prokopovBlessingInteractiveDevelopment2016}.
\end{quote}

These concerns are valid, but we need neither to spend time writing
complex assertions nor to perform formal checks in tests. With small
adjustments to test metadata, we can repurpose the test runner as a
tool for automatic code re-execution and value printing.

Later, such tests can be discarded or converted into standard
ones. Instead of explicitly setting metadata, we can configure these
values via HIDE's test runner interface or implement a separate
function for running tests in \texttt{inspect-nofailure} mode.

\begin{verbatim}
(test "small test"
 #:metadata '((inspect? . #t) (no-failure? . #t))
 (is #f)
 (is 'hi)
 (is some-variable-or-expression))
\end{verbatim}

Though empirically unverified, we considered this worth mentioning as
it may spark further ideas for readers.

\subsection{Rerunning Tests}
\label{sec:workflow-rerun}
\begin{itemize}
\item \textbf{Before}: 0.5-10 seconds (to evaluate the previous test
  expression).
\item \textbf{After}: sub seconds (to press rerun previous
  tests\footnote{Yes, and the time to actually evaluate the code of
    the test body, you are right :)}).
\end{itemize}

This workflow offers a streamlined alternative to navigating to a test
or test suite expression for manual evaluation (as demonstrated in
Section~\ref{sec:workflow-interactive-testing}). By leveraging HIDE's
interface and persisted test runner state, we re-execute the most
recent test suite with \texttt{C-c C-t C-t}. This eliminates
\textasciitilde10 seconds of context-switching between source files or
places while preserving critical focus on the primary development
task.

\subsection{Getting More Tests}
\label{sec:workflow-more-tests}
\begin{itemize}
\item \textbf{Before}: 0.5-10 seconds (to run a test suite).
\item \textbf{After}: 30 minutes (to sequentially run project tests).
\end{itemize}

So far we were using tests as an automated way for checking evaluation
results in HIDE/REPL.  This section formalizes a workflow for
incorporating these tests into the project, so we can further use them
for the quality assurance on CI or run locally to make sure everything
still integrates and works as expected or other purposes mentioned in
next sections.  The execution time of whole project test suite is
huge, but as we see soon, we rarely need to run it and when we do, we
do it asyncronously (Section~\ref{sec:workflow-asyncronous}).

With a bottom-up approach, we were experimenting and setting some
invariants using tests and updating corresponding function definitions
to match them. Now, when the experimentation phase is finished and
we've achieved desirable functionality, it's time to persist the
results. Actually, we are almost there. The function's code is already
written and probably even in the right file. The test is written also;
one just needs to move it into a corresponding test suite in a test
module (if it's not yet there)

If the project is not decided on the approach for organaizing and
persisting tests, we recommend to add a separate directory
\texttt{test/} to load-path and for every module
\texttt{src/our/module.scm} to have a corresponding
\texttt{test/our/module-test.scm} as described in
\autoref{sec:discovery}.

Producing and persisting tests allows for easier quality assurance and
fearlessly sharing work with colleagues, but leads to the same
10-minute run-time mentioned in \nameref{sec:problems}. Hang
tight—we're almost there. We need a comprehensive test suite for good
coverage, but we don't need to run all the tests every time.

\subsection{Asyncronous Test Execution or a Workaround}
\label{sec:workflow-asyncronous}
Before getting to the extremely efficient workflows, we need to
mention that there are cases when we need to run a big test suite (be
it a module's test suite or even the whole project test suite). One
such case is when we collect information about previous test runs
(which tests are failing, how much time a test takes to execute, and
so on) for further scheduling of efficient reruns
(\autoref{sec:workflow-rerun-failed}). While we consider this a setup
cost and not the primary part of the development workflow, we still
provide two tactics to mitigate the impact of long execution times.

HIDEs can run tests asyncronously and notify developer, when the
results are ready.  It's not a primary tool, more just a workaround
for rare cases we need it.  This allows developer to continue his
work, evaluate expressions, and do other things while tests are
running.

This of course can be useful in other scenarios, as can the second
tactic described in the next section.

\subsection{Parallel Test Execution}
\label{sec:workflow-parallel}
We optimize test execution time and decrease the delays, and there is
an obvious widely-useful optimization worth mentioning: parallel test
execution. Due to the design of the library (\autoref{sec:defining}),
tests are independent and can be executed in arbitrary orders,
including a parallel one (is it an order at all?). It provides faster
execution times: previously, 10 tests were executed in 5 seconds; now
they run in 0.8 seconds.

While it sounds simple—just run the tests on all the cores and get
faster execution—the implementation-wise, this functionality is
hard. Bringing the context of a failed test back into HIDE tooling
like a debugger is not complete yet and requires further work.  While
not yet integrating well with other workflows directly, the parallel
execution on its own already makes initial test run (needed for
collecting previous run information) times faster.

\subsection{Selecting Tests}
\label{sec:workflow-selecting}
\begin{itemize}
\item \textbf{Before}: 10 minutes (to evaluate the whole project test suite).
\item \textbf{After}: dozens of seconds (to run filtered tests).
\end{itemize}
The project has a comprehensive test suite, and running it takes 10
minutes. A developer can run it a few times a day, but due to the long
execution time, using it synchronously with uninterrupted development
flow is impossible.

The good part here is we rarely need to rerun the whole test suite
every time, and since we have the information about suite available in
our runtime, we can perform various operations on it to choose tests
most relevant to the current task. The simplest thing to do here is
filtering.

Advanced text editors like Emacs provide completion interfaces
(vertico+orderless, for example), which allow quickly filtering lists.
HIDEs like Ares/Arei utilize this completion interface to present a
list of loaded tests and it can be quickly filtered by a developer
using information about tests: test description, test suite, module or
the data from the previous run like execution time or fail/pass/error
status.

While we could implement CSS-like selectors or other complicated query
languages, a basic orderless matcher against all the stringified
information available about the test is already good enough. We don't
care about false positive matches because we will quickly get rid of
them by using workflows described in next two subsections.

The overall workflow looks like:
\begin{enumerate}
\item Load the whole project test suite using \nameref{sec:discovery}.
\item (Optionally) Run it once to collect last run information and
  enchance filtering capabilities.
\item Open filtering interface and type filtering query.
\item Run much narrower test suite in much less time.
\end{enumerate}

With this approach we can narrow it to a very particular test suite or
even a test, but often we are interested in a few test suites related
to the particular module and would like to work with them. In this
case filtering will bring us from dozens of minutes to dozens of
seconds, which is much better, but still far from subseconds.  Next
workflows explain how to get further in the execution time
optimization.

\subsection{Rerunning Failed Tests}
\label{sec:workflow-rerun-failed}
\begin{itemize}
\item \textbf{Before}: dozens of seconds (to evaluate the whole suite).
\item \textbf{After}: seconds (to rerun last failed tests).
\end{itemize}
Similiar to how we reduce the amount of tests to run in
\nameref{sec:workflow-selecting} by selecting only those relevant to
the current task, we can select tests that failed on the previous run.
One can do it manually with the same selection interface, but
alternatively we can use a flag described in
\nameref{sec:running-options}, and set it to rerun only tests that
failed during the previous run.  It has a special handy property: to
rerun the whole test suite if there is no failing tests from the
previous run left. Usually, it radically reduces the amount of tests
to run.  Thus, we are already quite close to a rapid feedback loop and
uninterrupted development flow here (but still not yet). However, a
few dozen of tests, even if they are quick can add up and lead to
seconds of delays. We will fix it in \nameref{sec:workflow-fail-fast},
but for now let's accept the delays and do the recap of the workflow:

\begin{enumerate}
\item Do the initial test run to obtain info about failing tests
\item Set rerun only failed tests flag
\item Rerun failed tests
\item Improve the code
\item Repeat step 3 until there is no failing tests left
\item Rerun all test from suite (it will automatically happen on the
  next rerun, when we get out of failing tests).
\end{enumerate}

Not rapid enough to reach a subsecond goal, but already quite quick
feedback loop.

% !!!!!!!!!!!!!!!!!!!!!!!!!!!!!!!!!!!!!!!!!!!!!!!!!!!!!!!!!!  An
% implementation note: Probably we can make a stack of previous run
% and pop out dynamically generated suites out of it.

\subsection{Fail Fast}
\label{sec:workflow-fail-fast}
\begin{itemize}
\item \textbf{Before}: seconds (to rerun last failed tests).
\item \textbf{After}: subseconds (to reach the first failing test)
\end{itemize}
During the development, we rarely care about how many tests are
failing at the moment. If there is at least one failing test, we want
to act on it. Fix it, go to the next one, so why bother waiting to
rerun the whole test suite? Just run the test suite until the first
failure occurs.  With the combination with previous workflow it means
that we often need to run only one test.

There is a place where the magic of workflows composability happens:
by combining a few not optimal but simple workflows, we get to the
point where the test execution requires the minimal possible amount of
time. We acknowledge that it may not always be desired to act on the
first failing test; we discuss an alternative approach allowing
similar gains but more flexible test ordering in
\nameref{sec:conclusion}.

While we have achieved subsecond test execution times, one critical
task remains: bringing debugging and inspection tools directly at the
point of test failure.

\subsection{Throw on Failure}
\label{sec:workflow-throw-on-failure}
\begin{itemize}
\item \textbf{Before}: minutes (to read the test report and setup the
  environment similiar to testing).
\item \textbf{After}: subseconds (just act)
\end{itemize}
What happens when the test fails? We usually start exploring why.
First of all, we look at the test report. When it has insufficient
information, we go deeper: look at what functions were in the stack at
the moment of failure and start exploring their source code. If this
doesn’t work, we try to set up the environment similar to what was in
the test to inspect values of variables and expressions.

All the time from the beginning of the \nameref{sec:workflows}
section, we were working with tests from the HIDE. We either directly
interacted with the test runner's API or through the UI provided by
HIDE. We had all the development and debugging tooling that entire
time at our disposal. To bring it to the context of test failure, we
just need to throw the exception on test failure and let the HIDE
handle the rest.  That can be done by setting corresponding option to
the test runner (see \autoref{sec:running-options}).

This will bring a stack trace viewer and interactive debugger, and we
will get immediate access to all the necessary clues: we can evaluate
arbitrary expressions in the lexical scope of the failure, see the
current state of the stack, and values of local variables.

Rapid test execution times, no more extra steps for digging into the
failure context, no more waste of time and focus.  What else we could
dream of?

\section{Related Work}
Our initial work was based on top of SRFI-64
\cite{bothnerSRFI64Scheme2006} and Geiser HIDE
\cite{ruizTopGeiserUser2009} we succesfully prototyped a few of the
workflows using them.  However, the SRFI-64 library was intended for
use with CLI, so it lacked runtime representation for tests and
delayed assertions/tests execution required for more advanced
workflows (\autoref{sec:workflow-selecting}).  We also took into
account the pitfalls \cite{volfStateSRFI64Wolfs2024} in the
specification and current implementation of SRFI-64, so we decided
it's reasonable to make a library from the scratch.  It allowed to
make test defining API simplier, reporting more versatile
(\autoref{sec:test-reporters}), and provided necessary runtime
entities, but led to much more complex test runner implementation.

The next huge influencer for our work is clojure.test
\cite{sierraClojuretestClojureV11212009}, we took the idea of
assertion macro \texttt{is} and reporters, started to implement
similiar API, but later diverged to address lacking runtime
representation for tests and delayed execution.  In addition to tests
we introduced test suites entitiy (inspired by JUnitXML
\cite{gurockJunitXMLFileFormat2023}) for clearer distinction of test
hierarchy organizational units.  Also, there is a very advanced Kaocha
test runner for clojure.test: it highlighted some weaknesses in
clojure.test test defining API, so we can avoid them in ours defining
API, and gave us inspiration for Fail Fast workflow
(\autoref{sec:workflow-fail-fast}).

According to development environments, CIDER
\cite{batsovCIDERClojureInteractive2012} gave us an impressive example
and better understanding of interactive test usage and integration of
a testing library with HIDE: it shows much reacher experience and
improved usability than usual text-based REPL. However, it is designed
for the use with Clojure, so we had to find something else for the
Scheme.  As we mentioned earlier, we started our initial prototype
with Geiser, it uses direct stdin/stdout integration with REPL, which
leads to blocking the rest of the functionality, when any operation is
in progress.  Later we had to switch to Ares/Arei
\cite{tropinAresAsyncronousReliable2023}\cite{tropinAreiAsynchronousReliable2023},
it based on asyncronous extendable nREPL protocol\footnote{the same
  protocol used in CIDER}, which allows to run tests asyncronously
without blocking the rest of the HIDE.

In addition to Clojure's ecosystem, we studied Common Lisp's one.  The
Practical Common Lisp \cite[Chapter~9]{seibelPracticalCOMMONLISP2005}
allowed us to kick off the implementation of the library.  The
extensive Comparison of Common Lisp Testing Frameworks
\cite{crolletonComparisonCommonLisp2021} showed the slight difference
in the functionality, but incompatible implementations, so we took it
into account and made API as flexible as possible, allowing the test
runner to alter the semantics of test definitions.  We believe it will
allow to avoid unecessary overdivirsity of testing solutions for
Scheme.

\section{Conclusion and Future Work}
\label{sec:conclusion}
% https://guidetogradschoolsurvival.wordpress.com/2011/04/15/how-to-write-future-workconclusions-2/
Our work fundamentally transforms testing from a separate interruptive
daunting task into a continuous rapid feedback mechanism integrated
into development workflows. By introducing runtime test entities and
fail-fast workflows, we achieve subsecond turnaround cycles and enable
immediate access to the HIDE's tooling for an extremely quick failure
investigation.  We hope this not only provides a theoretical
improvement of developer experience, but leads to real world
application and adoption by community. We expect it to influence
Scheme ecosystem in a positive way and enhance its quality.

To enable broader adoption, future work on SRFI standardization would
be beneficial. Although the library offers significant flexibility, we
remain uncertain whether all desired functionality—such as test
fixtures, expected failure handling, and similar features—could be
implemented on top of it without compromising usability. Therefore, we
plan to gather additional community feedback during the SRFI process
and potentially implement any identified gaps.

As mentioned in \autoref{sec:workflow-parallel}, there are more work
required for the reliable implementation of parallel test execution
and its results propogation to the rest of the HIDE tooling.

There are a few more workflow ideas that we have not implemented yet
but find reasonable.  The first one, called \texttt{Fail Fast, but
  Keep Running}, is a combination of
\nameref{sec:workflow-asyncronous} and
\nameref{sec:workflow-fail-fast}, we bring the HIDE debugging tooling
on the first failure, but keep running the tests, collecting
information for the next reruns and building a comprehensive report
for the developer in the background.  Similiar to
\nameref{sec:workflow-parallel}, it's hard implementation-wise, but at
the first glance looks promising.

The second one is called \texttt{Semi-Manual Scheduling}.  We can
provide a full-blown interface presenting all loaded test and allowing
different quick actions on them: like list sorting/filtering, or
particular entries moving/dropping.  That opens opportunities for more
ad-hoc workflows tailored for specific needs.  For example:
\texttt{Failing First Slow Last}, with a couple of key presses we can
sort tests by execution time and previous results and thus get the
list, where the first entries are more likely to produce a failure and
being the fastest to execute.

Also, test reloading remains a challenge. When editing application
code, respective test updates are often required. This necessitates
reloading modified tests and typically triggers a costly initial
execution of the test suite. We currently lack a robust solution for
this issue, though syntax tree diffing may provide a viable approach.

There is a great opportunity to improve even further on top of our
work, by utilizing language with content-addressable code
representation \cite{maziarzHashingModuloAlphaequivalence2021} and
highly-controlled side effects like Unison lang
\cite{chiusanoDocsUnisonTesting2013}.  It will allow to avoid
unecessary test re-executions, when the underlying code and its
dependencies didn't change. In combination with propogator networks
\cite{radulPropagationNetworksFlexible2009}
\cite{sussmanArtPropagator2009} it can make test reloading and
restarting seamless, reducing frictions, delays, and initial run costs
significantly.

Great thanks to all the people without whom this work won't be
possible.  [Add names and kudos after the review is finished].

%% Anna Prokhorova.

% %%
% %% The next two lines define the bibliography style to be used, and
% %% the bibliography file.
\bibliographystyle{ACM-Reference-Format}
\bibliography{bibliography}

\appendix
\section{Macro API}
\label{implementation}

The source code of the library will be attached verbatim in this
section? \cite{SuitblLibrarySource2025}

\end{document}